\documentstyle[eqsecnum,aps,twocolumn,epsfig,ilja1]{revtex}

\begin{document}
\draft
\title{Experimental generation of steering odd dark beams of finite length}
\author{A. Dreischuh, D. Neshev}
\address{Sofia University, Department of Quantum Electronics, 5, J.
Bourchier Blvd., BG-1164 Sofia, Bulgaria\\
(Fax.: +3592/9625276, E-mail: ald@phys.uni-sofia.bg)}
\author{G. G. Paulus$^1$, H. Walther$^{1,2}$}
\address{$^1$Max-Planck-Institut f\"ur Quantenoptik, Hans-Kopfermann-Stra{\ss}e
1, D-85748 Garching, Germany\\
(Fax.: +4989/329050, E-mail: ggp@mpq.mpg.de)}
\address{$^2$Ludwig-Maximillians-Universit\"at, Sektion Physik, Am Coulombwall 1,
D-85747 Garching, Germany\\
(Fax.: +4989/2891 4142, E-mail: Prof.H.Walther@mpq.mpg.de)}
\date{\today}
\maketitle

\begin{abstract}
In this work we report the first realization of odd dark beams of finite
length under controllable initial conditions. The mixed edge-screw phase
dislocations are obtained by reproducing
binary computer-generated holograms. Two effective ways to control the steering
of the beams are analyzed experimentally and compared with numerical
simulations.
\end{abstract}
%\pacs{{PACS:}42.65;42.65.Tg;42.65.Sf}
\pacs{{OCIS code:} 0190 4420; 0190 5940; 090 1760}
\preprint{HEP/123-qed}

\narrowtext

\section{Introduction}

\label{teil0} Physically, Dark Spatial Solitons (DSSs) are localized
intensity dips existing on stable background beams as a result of
an exact counterbalance of diffraction and nonlinearity. A
necessary condition for their existence is the presence of a phase
dislocation in the wavefront along which the phase is indeterminate
and the field amplitude is zero. Besides the intriguing physical
picture, the particular interest in the DSSs is motivated by their
ability to induce gradient optical waveguides in bulk
self-defocusing nonlinear media \cite{refc1,refc2,refc3,Yuri,Law}.
The only known truly two-dimensional (2D) DSSs are the Optical
Vortex Solitons (OVSs) \cite{refc2}, whereas in one transverse spatial
dimension DSSs manifest themselves as dark stripes
\cite{refc4}. The odd initial conditions required to generate a
fundamental 1D DSS correspond to an abrupt $\pi$-phase jump
centered along the irradiance minimum of the stripe. The OVSs have
a more complicated phase profile described by $\exp(im\varphi)$,
where $\varphi$ is the azimuthal coordinate in a plane
perpendicular to the background beam propagation direction and $m$
-- the so-called topological charge (TC) -- is an integer number.
This phase function ensures a $\pi$-phase jump in each diametrical
slice through the vortex core. Fundamental DSSs of these types have
the common feature of zero transverse velocity if no
perturbations are present. In contrast to that, ring dark solitary
waves \cite{refc5} slowly change their parameters even when born
from ideal odd initial conditions \cite{refc6}.

In the pioneering work of Nye and Berry \cite{refc7} it is conjectured that
mixed edge-screw dislocations cannot exist. Despite that, almost two decades
later an indication for their existence was found \cite{refc8} for two
interacting optical vortices of opposite topological charges. In our recent
experiments on the generation of quasi-2D DSSs \cite{refc9} we found that
moderate saturation of the nonlinearity can stabilize the snake
instability which usually leads to their decay.
This made possible the first identification of 1D Odd Dark Beams
(ODBs) of finite length with their characteristic edge-screw phase
dislocations \cite{refc9,refc10}. The mixed dislocation forces the dark
beams to steer in space. This appears to be of practical interest \cite{refc10}
provided that there are effective ways to control the ODBs transverse velocity.

In this article we report the first experimental realization of steering odd
dark beams of finite length with mixed phase dislocations under controllable
initial conditions. Two approaches to control their transverse velocity are
investigated experimentally and compared with numerical simulations.

\section{Experimental setup and results}
\label{teil1}

\subsection{Computer-generated holograms}
\label{teil1.1} The phase portrait of the mixed edge-screw dislocation (see
Fig.~1 in Ref.\cite{refc10}) consists of a pair of semi-helices with a phase
difference of $\pi$ to which an effective topological charge of $\pm 1/2$
can be ascribed. Their spatial offset $2b$ determines the length of the
edge part of the dislocation and ensures a phase jump of $\Delta \varphi$ in
the direction perpendicular to the dark stripe of finite length.
The phase function of this mixed phase dislocation can be described by
\begin{equation}
\label{gleich1}\Phi_{\alpha,\beta}(x,y)=
\Delta \varphi\left\{ -\frac{\beta}{\pi}\arctan\left( \frac{\alpha x}{%
%y+b\beta} \right) +\frac{1}{2}(1-\alpha)\text{sgn}(x)\right\} ,
y+b\beta} \right) +\frac{(1-\alpha)}{2}\text{sgn}(x)\right\} ,
\end{equation}
where $x$ and $y$ denote the transverse Cartesian coordinates
perpendicular and parallel to the dark beam. $2b$ stands for the
length of the edge part of the dislocation and
\begin{equation}
\label{gleich2}\alpha= \left\{
\begin{array}{ll}
0 & \text{for } |y|\leq b \\ 1\text{ and }\beta=-1 & \text{for }y> b \\
1 \text{ and } \beta=1 & \text{for } y\le -b
\end{array}
\right. .
\end{equation}

%\begin{equation}
%\label{gleich2}
%\alpha= \left\{ \matrix {
%0&\mbox{for $|y|\leq b$} \cr
%1 \mbox{ and $\beta=-1$} & \mbox{for $y > b$} \cr
%1 \mbox{ and $\beta=1$} & \mbox{for  $y\le -b$}  }
%\right. .
%\end{equation}

The pattern of the Computer-Generated Holograms (CGHs) used to produce this
phase distribution consists of parallel lines which become
curved at the position of the semi-vortex cores.
In the edge part of the dislocation
they terminate and reappear shifted, for a $\pi$-jump by one half
of the pattern period.
Holograms with such structures correspond to interference
lines shifted along an imaginary line of finite length and to curved
lines limiting the dislocations as observed in our previous experiment (see
Fig.~8 in \cite{refc9}). The binary CGHs used are photolithographically
fabricated with a grating period of $20\mu m$. Several holograms with various
lengths of the edge part of the dislocation are etched on a common substrate.
Special attention was paid to align
the edge parts of the different dislocations correctly on their common
substrate. The simplicity of varying the dislocation length and magnitude of
the phase jump is the main advantage \cite{refc11} of the approach we have
chosen. The diffraction
efficiency at first orders is measured to be $9\%$, close to the theoretical
$10\%$ limit for binary holograms.
The unavoidable quantization inaccuracy of $\pi/24$ \cite{refc12} for holograms
of this type is negligible for measurements with phase jumps of
$\Delta\varphi=3\pi /4$, $\pi$, and $5\pi/4$ which are presented in this work.

\subsection{Experimental setup}
\label{teil1.2} The setup used is similar to that in our previous
experiments (see Fig.~1 in Ref.\cite{refc9}). Briefly, the beam of a
single-line $Ar^{+}$ laser ($\lambda =488nm$) is used to reconstruct the CGHs.
The first-diffraction-order beam with the phase dislocation nested in is
filtered through a slit and is focused on the entrance of the $10cm$ long
Nonlinear Medium (NLM). After the desired propagation path length,
the beam is deflected by a prism immersed in the nonlinear liquid and
is projected directly on a Charge-Coupled Device (CCD) camera array with a
resolution of $13\mu m$. The NLM is ethylene glycol dyed with DODCI
(Lambdachrome) to reach an absorption coefficient of $0.107cm^{-1}$. In a
calibration measurement we generated 1D DSSs by using CGHs of the type
described in Sec.~\ref{teil0}. The soliton constant $Ia^2$ (i.e. the product of
the background-beam intensity $I$ and the square of the dark beam width $a$
measured at the $1/e$-level) was found to reach its asymptotically constant
value for input powers of $P_{\rm sol}^{1D}\approx 33mW$. It is known that
thermally self-defocusing liquids are both nonlocal and saturable. Since the
saturation of the nonlinearity is able to modify the ODBs transverse
velocity and profile, we needed to estimate it and to account for it
in our numerical calculations. In an independent measurement we realized a
self-bending scheme similar to that used in \cite{refc13,refc14}. The
asymmetry required was introduced by an intentional tilt of the prism
immersed in the NLM, which resulted in different nonlinear propagation
path-lengths for the different parts of the background beam. The strength of
the self-bending effect was measured in the near field. For an absorptive
nonlocal medium the choice of a suitable saturation model is not trivial
(\cite{refc15}, see also Sec.~IV of \cite{refc16}). We found a good fit
for the experimental data with the equation $\Delta y\sim
I/(1+I/I_{sat})^\gamma$.
Using it we estimated $P_{sat}\approx 100mW$ and $\gamma =3$. In addition
to the careful alignment of the CGHs on the substrate, the holograms were
reproduced to achieve {\it vertical} dark beam steering, which is not sensitive
to possible undesired weak horizontal self-deflection of the background
beam. Changing the ODB parameters (length-to-width ratio and magnitude
of the phase jump) was performed by a strict horizontal translation of the
substrate. The accuracy of the alignment was tested by checking for equal
steering of the ODBs reproduced from two identical holograms placed at opposite
ends of the series of aligned CGHs. In this work the ODBs are identified by the
corresponding lengths $2b$ of the edge portions of the dislocations in CGH
pixels ($1 pix.=5 \mu m$) as encoded in the holograms. The deflection
$\Delta x$ of the dark beams is measured in units of CCD camera pixels.
We estimate that measurements with an encoded dislocation length of $b/5 \mu m$
CGH pixels corresponds to a dislocation
length-to-ODB width ratio $(2b/a)= 1/10 b/5 \mu m$ in the numerical simulations
(e.g. 14 pix. dislocation length corresponds to $b/a=1.4$ in the simulations).

\subsection{ODB steering vs. dislocation length}
\label{teil1.3} All data presented in this subsection refer to $\pi$ phase
jumps across the edge parts of the mixed dislocations. In Fig.~1a we plot
the deflection $\Delta x$ of ODBs with different dislocation lengths for input
powers of $1.7mW$ (circles), $33mW$ (triangles) and $67mW$ (squares). The
data obtained at $P=1.7mW$ refer to a linear regime of propagation. The
results at $33mW$ and $67mW$ are extracted from the experimental pictures
shown in Fig.~1b and Fig.~1c, respectively, which are recorded for a
nonlinear propagation path length of $z=8.5cm$. The general tendency of
linear increase of the deflection with decreasing the dislocation length is
clearly expressed (Fig.~1a, solid line). The shortest mixed phase dislocation
encoded was only $1 pix.$ long. In this case the strong deviation from the
linear dependence is caused by the annihilation of the semi-charges due to
the shortening of the edge part of the dislocation. It will be shown later
that this shortening accelerates for higher input powers (intensities). For
that reason, even the ODB with an initially $10 pix.$ long phase dislocation
appears gradually less deflected at $P=67mW$ as compared to the case of
$P=33mW$ (Fig.~1a). In Fig.~1b,c the thick solid lines are
intended to denote the positions of the ODBs at the entrance of the NLM.
Because ODB steering is present also in the linear regime of propagation
(Fig.~1a, dots) these positions (with respect to the dark beam intensity
minimum and trailing peak maximum) are identified by numerical simulations
for $b/a=2.5$. The identification corresponds to an encoded dislocation
length of $25 pix.$, i.e. to the most right frame shown in Fig.~1b,c.
Somewhat surprising is the weak sensitivity of the ODB
deflection vs. background beam power (intensity) far from TC annihilation.
It can be intuitively understood by recalling the known interaction scenario
of well-separated OVSs of opposite TCs \cite{refc17,refc18}. In this case
the attraction between the OVSs is negligible as compared to their
translation as a pair. At a constant input power of $33mW$ we measured the
ODB deflection vs. nonlinear propagation path length (Fig.~2). As expected,
the ODB with $10 pix.$ long dislocation has higher transverse velocity
than that one with $22 pix.$ long dislocation.
The linearity in the dependencies is also well pronounced.

\begin{figure}
\centerline{\epsfig{file=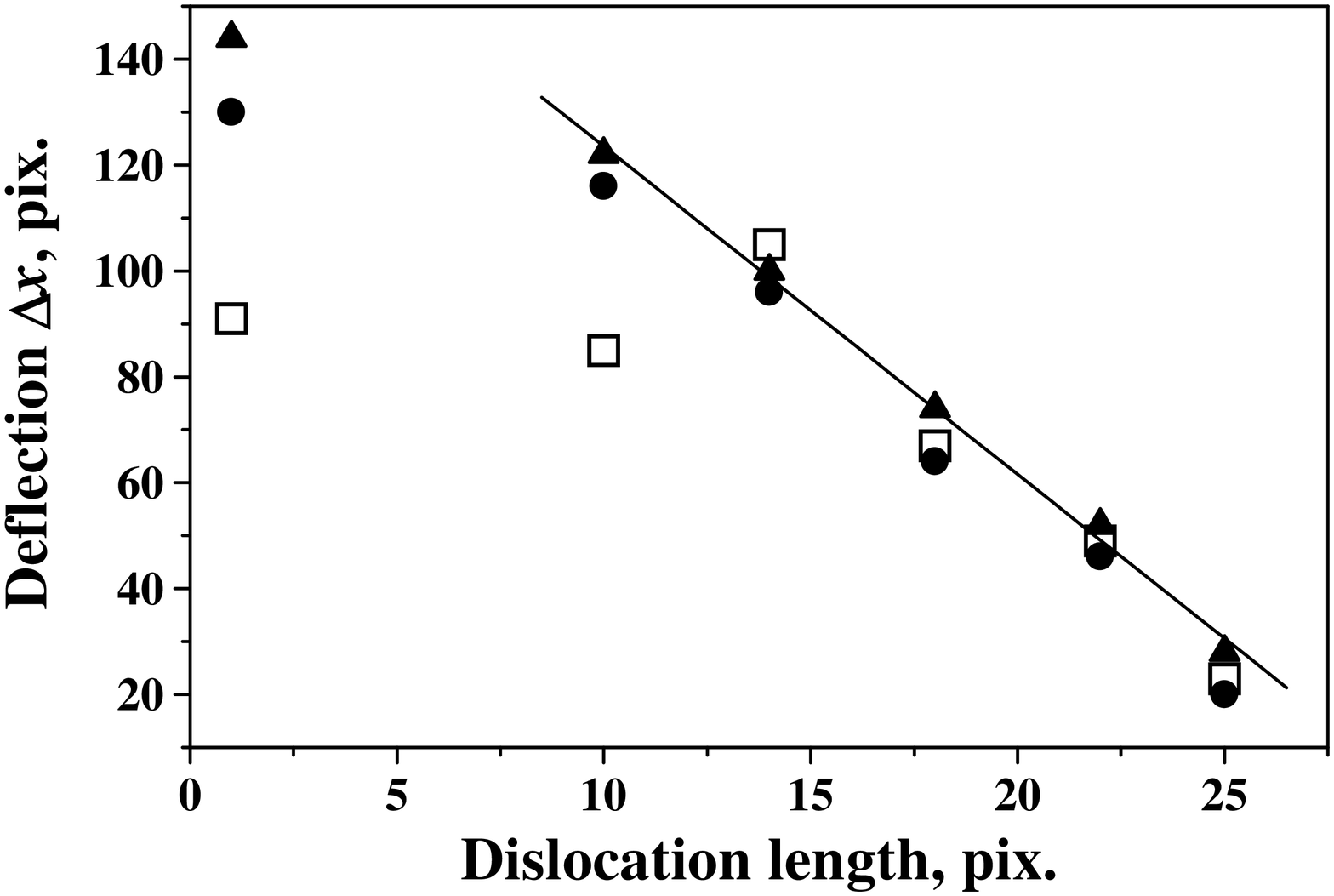,width=3in,clip=} }
\centerline{\epsfig{file=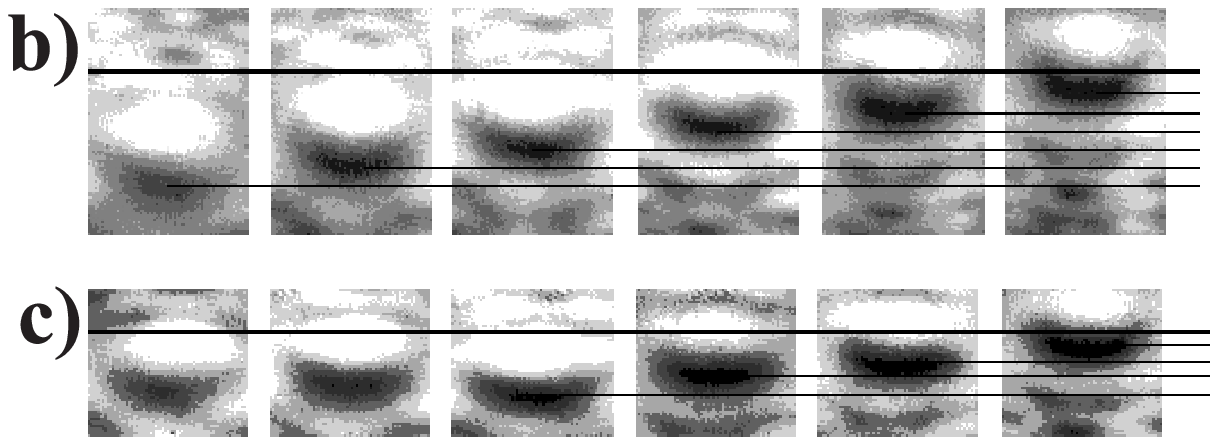,width=3.2in,clip=} }
\caption{Deflection of odd dark beams of finite length vs. dislocation length
at input powers of $1.7mW$ (dots), $33mW$ (triangles), and $67mW$
(squares) for $z=8.5cm$ and $\Delta \varphi =\pi$ (a). The points are extracted from
the frames shown in Fig.1b ($33mW$) and Fig.1c ($67mW$). From left to right the
frames correspond to
$b=1, 10,14,18,22,$ and $25pix.$, respectively. The thick solid line indicates the
calculated
position of the ODB at the entrance of the NLM for $b=25pix.$.}
\label{p6fig1}
\end{figure}

\begin{figure}
\centerline{\epsfig{file=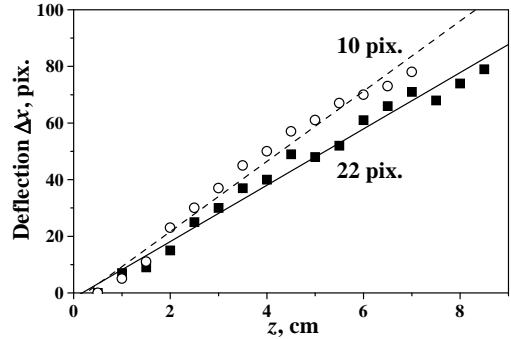,width=3in,clip=} }
\caption{Deflection $\Delta x$ vs. nonlinear propagation path length $z$ for
encoded dislocation lengths of $10pix.$ (circles) and $22pix.$ (squares).
The dashed and the solid lines are the respective linear fits. ($P=33mW$; $\Delta \varphi =
\pi$).}
\label{p6fig2}
\end{figure}

\subsection{Phase control of the ODB steering}
\label{teil1.4} As a second possible way to control the dark beam deflection,
we considered the variations in the magnitude of the phase jump
$\Delta \varphi$ across the edge part of the mixed dislocation. In Fig.~3 we
compare the experimental dependencies
$\Delta x(\Delta \varphi )\mid _{b=22pix.}$ (squares) and
$\Delta x(b)\mid _{\Delta \varphi =\pi }$ (dots). The straight lines are the
respective linear fits. Because a problem in encoding
a larger set of phase jumps in the CGHs was recognized too late, we measured
the deflection $\Delta x$ at $\Delta \varphi =3\pi /4$, $\pi $, and $5\pi /4$
only. In view of that the linear fit of the phase dependence in Fig.~3 appears
to be assailable. Its linearity, however, is confirmed by numerical simulations
(see Sec.~\ref{teil2}). The dependence of $\Delta x$ on $\Delta x(\Delta
\varphi )$ and $\Delta x(b)$ has been plotted in one figure in order to
underline the fact that it appears to be easier to deflect the ODB
by controlling the phase than by controlling  the dislocation length. The
measurements are performed at a constant power of $33mW$.

\begin{figure}
\centerline{\epsfig{file=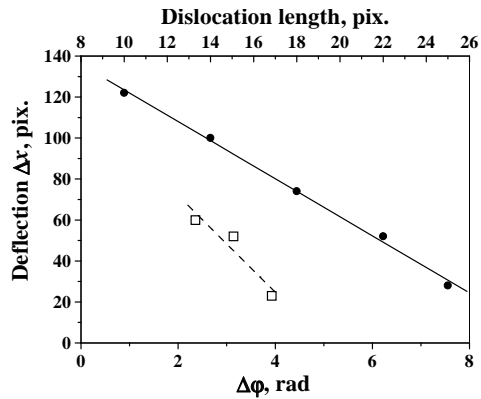,width=3in,clip=} }
\caption{ODB deflection vs. magnitude of the phase jump $\Delta\varphi$
(hollow squares) and the dislocation length as encoded in the CGHs (dots).
The straight lines are linear fits. ($P=33mW$).}
\label{p6fig3}
\end{figure}

\subsection{Power/Intensity dependencies}
\label{teil1.5}
The ability of the dark spatial solitons \cite{refc2,refc3,Yuri,Law} (and the
dark spatial waves \cite{refc19,refc20}) to induce gradient all-optical
waveguides in bulk self-defocusing NLM originates in the negative nonlinear
correction to the linear refractive indices of the media. In view of that, the
intensity dependencies remain undoubtfully of interest, despite of the low
sensitivity of the ODB steering on the input power (intensity).
In Fig.~4 we present experimental data on the power dependence of the
length of the edge portion of the mixed dislocation for $\Delta \varphi =\pi$
and for two different lengths of $14pix.$ and $22pix.$ encoded in the
holograms. It is easy to understand that the (mixed) phase dislocations do not
remain sharp and of an unchanged magnitude when the (odd) dark beam steers
\cite{refc21}. The dislocation lengths are estimated by evaluating the
respective longitudinal ODB slices at $5\%$ of the background beam intensity
(i.e. at the actual noise level in the recorded frames). Generally, the
dislocation length decreases monotonically with increasing the input power.
Asymptotically, the dislocation flattens and disappears, provided that the
ratio $b/a$ is less than 2. As mentioned in \cite{refc10}, at $b/a\sim 4$ the
ODBs should be expected to bend due to the snake instability \cite{refc15}. In
fact we observed such a behavior for ODBs with encoded dislocation lengths
ranging from 4 to 6. A vortex-beam creation is recognized by the convergence of
two neighboring interference lines in one. However, the vortices formed by
this instability remained with highly overlapping cores
\cite{refc15,refc22}.

\begin{figure}
\centerline{\epsfig{file=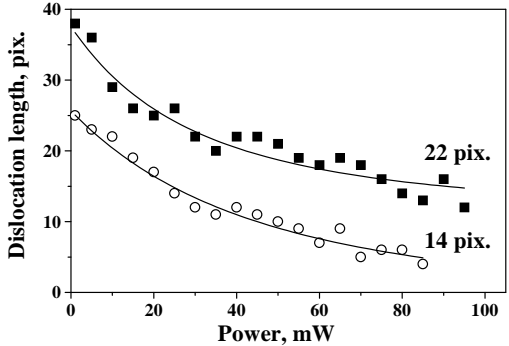,width=3in,clip=} }
\caption{Edge dislocation length vs. input background beam power for two
different dislocation lengths encoded in the CGHs.
($\Delta\varphi=\pi; z=8.5cm$)}
\label{p6fig4}
\end{figure}

\begin{figure}
\centerline{\epsfig{file=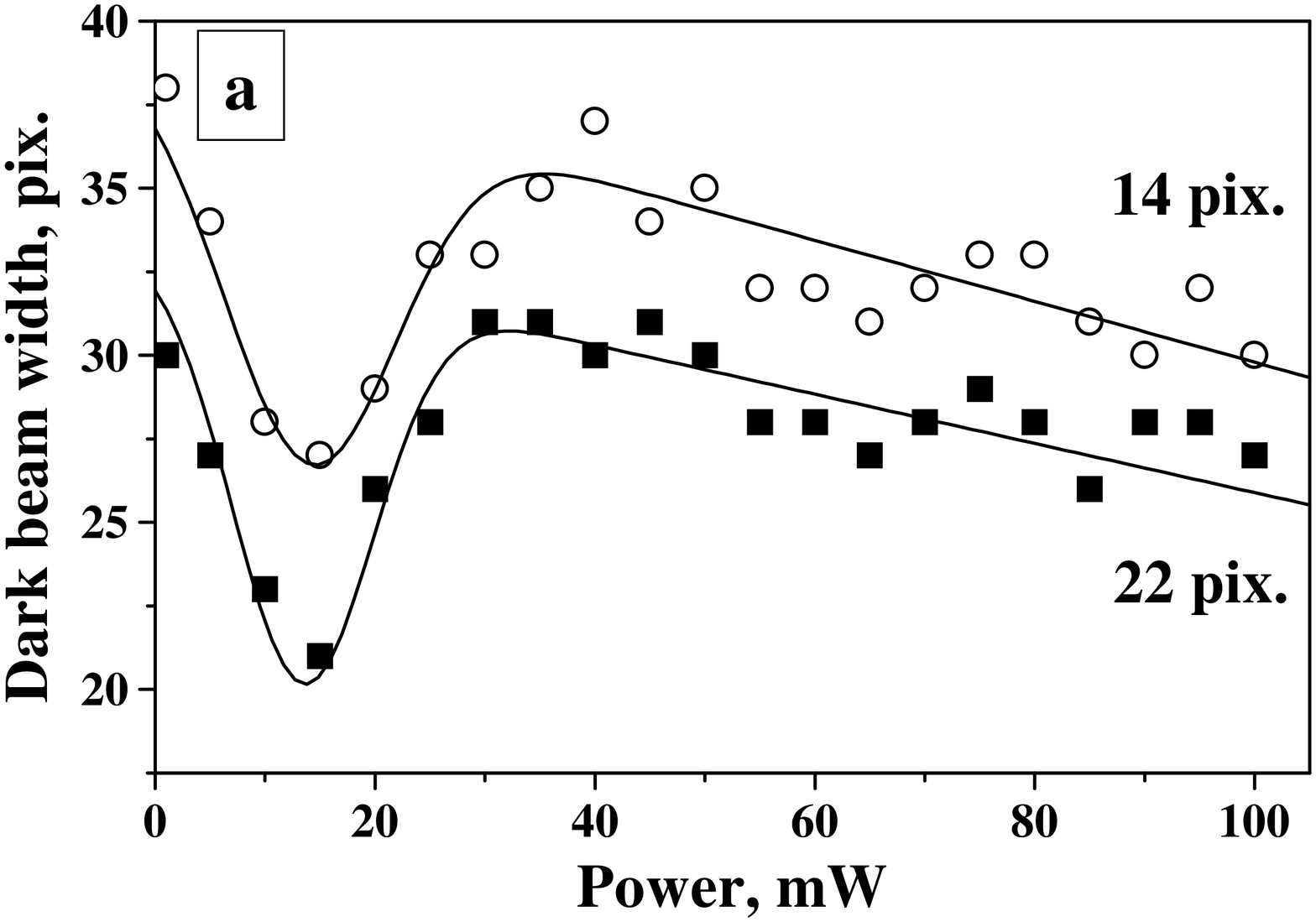,width=3in,clip=} }
\centerline{\epsfig{file=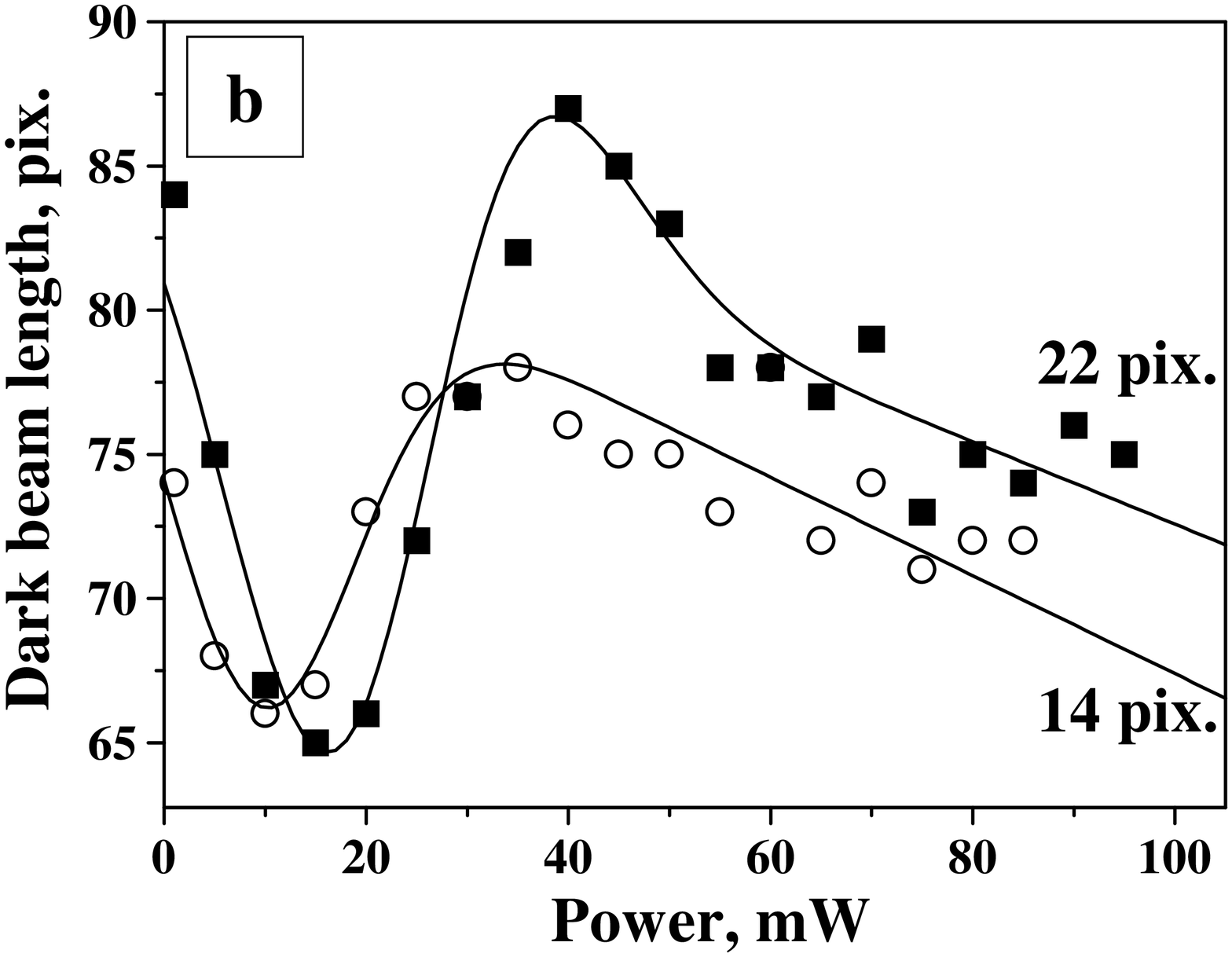,width=3in,clip=} }
\caption{ODB width (a) and length (b) (at the $1/e$ level) vs. input
background-beam power for two different dislocation lengths encoded in the
CGHs. ($z=8.5cm$)}
\label{p6fig5}
\end{figure}

In Fig.~5a,b we plot the measured ODB widths (a) and their lengths (b) at the
$1/e$-level as a function of the background beam power. The initial lengths of
the mixed phase dislocations are denoted in pixels as encoded on the respective
CGHs, whereas the widths and lengths at the exit of the NLM are measured in
units of CCD camera pixels. The strong decrease in both the ODB width and
length up to $17 mW$ (approximately $P_{sol}^{1D}/2$) is followed by an
approximate recovering of the width and length at approximately $P_{sol}^{1D}$.
At higher input powers both transverse quantities decrease but the tendency is
slower as compared to the situation
below $P_{sol}^{1D}/2$ and stabilizes asymptotically at high saturation. As it
will be discussed in the next section, the minima in the curves plotted in
Fig.~5a,b result from the reshaping of the OD beams
at the particular $1/e$ intensity-level chosen for evaluation.
The estimation there shows that the
first rapid decrease in both transverse dimensions of the ODBs does not
correspond to a `soliton constant' formation. Generally speaking,
the ODBs analyzed are no solitary waves in the widely adopted sense, since
they do not survive, for instance, a collision with a second ODB steering
in the opposite direction \cite{refc10}. Nevertheless, the narrowing in both
transverse directions for higher powers (intensities) should improve their
guiding ability when signal beams or pulses are to be transmitted inside the
ODBs and deflected in space. This feature will be addressed elsewhere.
It is interesting to note that, independent of the length of the mixed
dislocations ($2b$), the widths ($a$) of the dislocations should be initially
equal (in the near field behind the CGHs), but appeared different near
the entrance of the NLM. The estimation has shown a CGH-to-NLM distance of
approximately 4 Rayleigh diffraction lengths with respect to the initial ODB
width. The well pronounced separation between the curves in Fig.~5a should be
attributed to the different two-dimensional diffraction at different initial
ODB length-to-width ratios.

\section{Numerical simulations}
\label{teil2}
In our numerical simulations we tried to model the experimentally obtained
dependencies by accounting for the estimated moderate saturation of the
nonlinearity ($P_{sol}^{1D}=33mW$; $P_{sat}=100mW$ at $\gamma =3$; see
Sec.~\ref{teil1.2}). The $(2+1)$-dimensional nonlinear evolution of the
steering ODB of finite length in the bulk homogeneous and isotropic NLM is
described by the generalized nonlinear Schr\"odinger equation
\begin{equation}
\label{Schroedinger}i\frac{\partial E}{\partial\zeta}+\frac{1}{2} \left(
\frac{ \partial^2}{\partial\xi^2}+ \frac{\partial^2}{%
\partial\eta^2}\right) E-\frac{L_{Diff}}{L_{NL}}
{|E|^2\over (1+s|E|^2)^\gamma}E=0 ,
\end{equation}
where the transverse spatial coordinates are normalized to the initial
dark-beam width $(\zeta=x/a,\eta=y/a)$, and the propagation path length is
expressed in Rayleigh diffraction lengths $L_{\rm Diff} =ka^2 $. Further,
$L_{NL}=(k n_2 I_0)^{-1}$ is the nonlinear length, $k$ is the wavenumber inside
the NLM, and $I_0$ is the background beam intensity.
The adopted correction for the nonlinear refractive index is
\begin{equation}
\label{gleich4} \Delta n=n_2|E|^2/(1+s|E|^2)^\gamma,
\end{equation}
with $s=P_{sol}^{1D}/P_{sat}=0.3$. As it was done in \cite{refc10}, the slowly
varying electric-field amplitude of the ODB was chosen $tanh$-shaped
\begin{equation}
\label{gleich5}E(x,y)=\sqrt{I_0}B(r_{1,0}(x,y))\tanh\left[ \frac{%
r_{\alpha,\beta}(x,y)}{a}\right] e^{i\Phi_{\alpha,\beta}(x,y)},
\end{equation}
where $r_{\alpha,\beta}(x,y)=\sqrt{x^2+\alpha(y+\beta b)^2}$ is the
effective Cartesian/radial coordinate, $\Phi_{\alpha,\beta}$ is the phase
distribution of the mixed edge-screw dislocation (see Eq.~\ref{gleich1}), and
$\alpha$ and $\beta$ are given by Eq.~\ref{gleich2}. The width $w$ of the
super-Gaussian background beam
\begin{equation}
\label{gleich6}
B(r)=\exp\left\{ -\left(
\sqrt{\frac{x^2+y^2}{w^2}}\right)^{14}\right\}
\end{equation}
is chosen to exceed at least 15 times the ODB lengths. The model equation
(\ref{Schroedinger})
was solved numerically by the beam propagation method on a $1024\times 1024$
grid. It should be mentioned, that the initial width $a(z=0)$ of the ODB of
finite length was chosen to correspond to that of an infinite 1D ODSS
($a=a_{sol}^{1D}=const.I_0^{-1/2}$). It was proven numerically that the ODB
deflection is insensitive to the particular value of $a$, and Figs.~6-8 are
generated under this assumption. Actually, the nonlinearity causes an
appreciable reshaping of the beams, in particular in the first evolution stage
when the ODB starts steering, see Fig.~2 in Ref.~\cite{refc10}). In order to
improve the similarity between the experimental data (Figs.~4 and 5) and the
numerical results (Figs.~9 and 10), an initial ODB width twice as large as in
the experiment is assumed. In Fig.~6 we plot the ODB deflection vs.
$b/a \mid _{z=0}$ for different input powers (intensities). All data presented
in this section refer to a normalized propagation path length of $z=4L_{NL}$,
which corresponds to that estimated for the experiment. Nevertheless all
calculations are carried out up to $10L_{NL}$ whereby no qualitative deviation
from the tendencies discussed is seen. The linearity in the ODB deflection vs.
$b/a$ is well obeyed except for $b/a>2.2$. The longer ODBs steer slower, bend
slightly, and decay into pairs of vortex beams for $b/a>4$.
In the linear regime of propagation, the ODBs also deflect but the deflection
is stronger at higher input powers/intensities. This is more pronounced for
shorter dislocations. In the experimental data (Fig.~1a), this behavior is much
weaker, rather the deflection remains within the experimental accuracy. Looking
for an adequate explanation, in a series of simulations we checked that a
$\pm30\%$ inaccuracy in estimating $P_{sat}$ results only in $\pm5\%$ deviation
in the ODB deflection at $z=4L_{NL}$. The observed tendency of a decrease
of the ODB steering velocities at increased saturation is well understood
\cite{refc23} but seems insufficient in quantity. We attribute the
absence of a well expressed power dependence in Fig.~1a to the NLM nonlocality.
In a separate experiment it was estimated, that the nonlocality in this medium
is negligible on a spatial scale of several hundred of micrometers only
\cite{refc16}.

\begin{figure}
\centerline{\epsfig{file=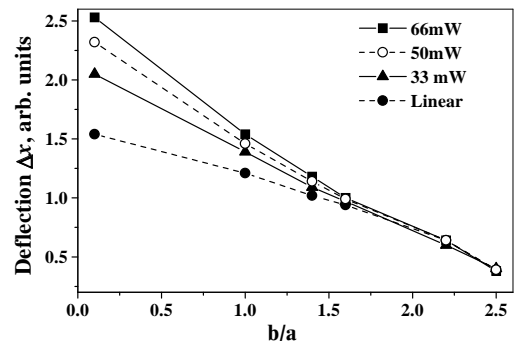,width=3in,clip=} }
\caption{Calculated ODB steering vs. $b/a$ for different input powers with
$\Delta \varphi = \pi$.}
\label{p6fig6}
\end{figure}

\begin{figure}
\centerline{\epsfig{file=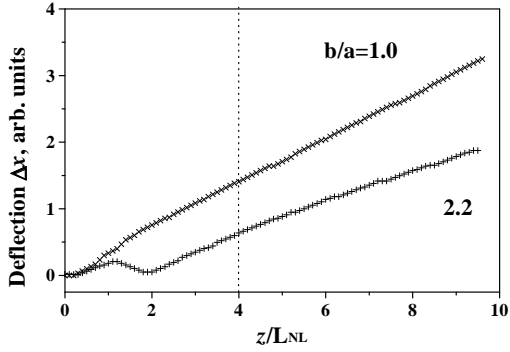,width=3in,clip=} }
\caption{ODB steering along the NLM for $b/a=1.0$ and $2.2$.
Crosses - some $10\%$ of the numerical data. Vertical
dashed line - propagation distance $z/L_{NL}=4$ corresponding to the
experimental conditions at $33mW$.}
\label{p6fig7}
\end{figure}

In Fig.~7 we plot the ODB deflection vs. the nonlinear propagation path length
$z/L_{NL}$ for $b/a=1.0$ and $2.2$. As in the previous figure the magnitude of
the edge part of the phase dislocation is set to $\Delta\varphi =\pi$. Ones the
ODB starts steering, its transverse velocity
remains constant (see Fig.~2). The longer ODBs with longer edge dislocations,
however, emit dispersive waves in their first evolution stage ($z<1L_{NL}$).
This causes a `delay' in the steering along the NLM (Fig.~7, lower curve).

\begin{figure}
\centerline{\epsfig{file=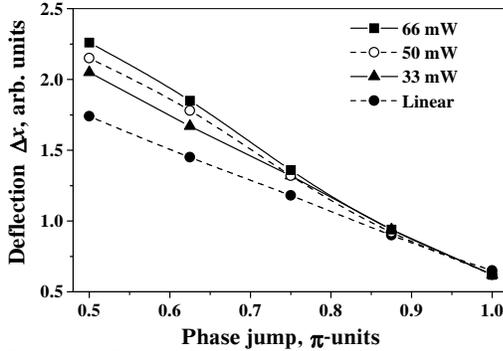,width=3in,clip=} }
\caption{Deflection of ODBs vs. phase jump $\Delta \varphi$ for $b/a=2.2$ and for
different input powers.}
\label{p6fig8}
\end{figure}

In Fig.~8 we present results obtained for the phase-dependent control of the
ODB deflection at different input powers (intensities). In qualitative
agreement with the experimental observation, the linear increase in the ODB
steering with decreasing the magnitude of the phase jump down to
$\Delta \varphi =0.5\pi$ is evident. The comparison of Figs.~6 and 8 confirms
the conclusion that at a fixed nonlinear propagation distance the
phase-controlled ODB deflection is more efficient as compared to that by
varying the $b/a$ ratio.

\begin{figure}
\centerline{\epsfig{file=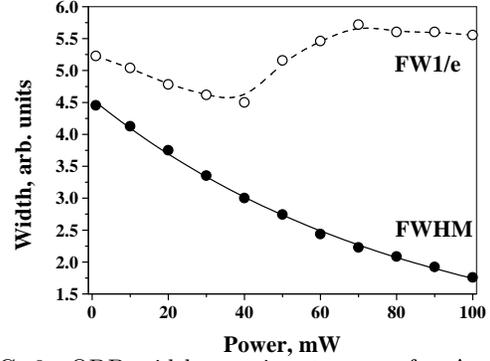,width=3in,clip=} }
\caption{ODB width vs. input power for $\Delta \varphi = \pi$ and $b/a=1.4$.
(Dashed curve: full width at the $1/e$-intensity level; solid: FWHM).}
\label{p6fig9}
\end{figure}

Fig.~9 is intended to clarify the origin of the non-monotonic power
dependencies of the ODB widths (and lengths) observed at input powers below
$P_{sol}^{1D}$ (see Fig.~5). The solid line represents the ODB full width at
half maximum (FWHM), the dashed one the full width at the $1/e$-intensity
level. In these simulations $\Delta \varphi =\pi$ and $b/a=1.4$ are
assumed. Qualitatively, we obtained the same curves also for
$b/a=2.2$ by accounting for the initial free-space propagation in
the experiment (from the CGH to the entrance of the NLM; $z\approx
4L_{Diff}$). The minimum in the ODB width evaluated at the $1/e$ level
originates from the reshaping of the beam profile which is caused
by the moderate saturation. A similar reshaping is reported in
\cite{refc24} (see Figs.~2-4 therein). At $b/a=2.2$ the data
obtained for the ODB length vs. input power (intensity) were found
to be even more sensitive to the intensity level of evaluation.
Because of the transverse steering of the ODBs of finite length the edge
portions of the dislocations shorten monotonically with increasing the
background beam power (intensity) (Fig.~10). The numerical results are in a
very good qualitative agreement with the experimental ones (Fig.~4).

\begin{figure}
\centerline{\epsfig{file=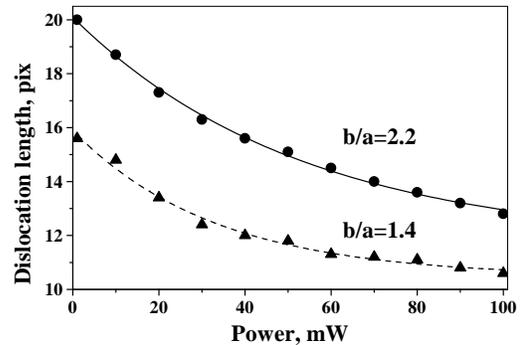,width=3in,clip=} }
\caption{Length of the edge portion of the mixed phase dislocation vs. input
power for $\Delta \varphi \mid_{z=0} =\pi$ and $b/a=1.4$ and $2.2$,
respectively.}
\label{p6fig10}
\end{figure}

\section{Conclusion}
The results presented show that the inherent steering dynamics of odd dark
beams of finite length can be effectively controlled by varying both the
magnitude $\Delta \varphi$ and the relative length $b/a$ of the mixed
edge-screw phase dislocation. The background-beam intensity has weak influence
on the steering but is important for keeping the optically-induced gradient
waveguides steep, which is crucial for all-optical guiding, deflection and
switching of signal beams or pulses. Since the mixed phase dislocations shorten
and flatten along the nonlinear media (tending asymptotically to washout) the
ODBs seem to be promising primarily for future short-range all-optical
switching devices.

\acknowledgments
A. D. would like to thank the Alexander von Humboldt Foundation for the
award of a fellowship and the opportunity to work in the stimulating
atmosphere of the Max-Planck-Institut f\"ur Quantenoptik (Garching,
Germany). This work was also supported by the National Science Foundation of
Bulgaria.

\end{document}